\documentstyle[epsfig,a4p,12pt,rotating]{article}
\newcommand {\Hpm}           {\mbox{${\mathrm H}^\pm$}}
\newcommand {\Hp}            {\mbox{${\mathrm H}^+$}}
\newcommand {\Hm}            {\mbox{${\mathrm H}^-$}}
\newcommand {\HH}            {\Hp\Hm}
\newcommand{\qqp}{\mbox{$\mathrm{q\overline{q}^\prime}$}}
\newcommand{\qpq}{\mbox{$\mathrm{q^\prime\overline{q}}$}}
\newcommand{\qppqppp}{\mbox{$\mathrm{q^{\prime\prime}\overline{q}^{\prime\prime\prime}}$}}
\newcommand {\qq}            {\mbox{${\mathrm q} \bar{\mathrm q}$}}

\newcommand {\ee}            {\mbox{${\mathrm e}^+{\mathrm e}^-$}}
\newcommand {\mumu}          {\mbox{${\mu^+\mu^-}$}}

\newcommand {\tautau}        {\mbox{${\tau^+\tau^-}$}}

\newcommand {\tpnu}          {\mbox{${\tau^+\nu_{\tau}}$}}
\newcommand {\tmnu}          {\mbox{${\tau^-{\bar{\nu}}_{\tau}}$}}

\newcommand {\W}             {\mbox{W$^{\pm}$}}
\newcommand {\WW}            {\mbox{W$^+$W$^-$}}

\newcommand {\mW}            {\mbox{$M_{{\mathrm W}^{\pm}}$}}
\newcommand {\mH}            {\mbox{$M_{\mathrm{H}^\pm}$}}
\newcommand {\mA}            {\mbox{$M_{\mathrm{A}^0}$}}

\newcommand {\csbar}         {\mbox{${\mathrm c} \bar {\mathrm s}$}}
\newcommand {\cbars}         {\mbox{${\mathrm s} \bar {\mathrm c} $}}
\newcommand {\cbbar}         {\mbox{${\mathrm c} \bar {\mathrm b}$}}
\newcommand {\etal}          {et al.}

\newcommand {\Ecm}           {\mbox{$\sqrt{s}$}}

\newcommand{\sseven}     {\mbox{$\sqrt{s}$~=~172 GeV}}
\newcommand{\sthree}     {\mbox{$\sqrt{s}$~=~130$-$136 GeV}}
\newcommand{\roots}     {\sqrt{s}}
\newcommand{\wpair}{\mbox{$\mathrm{W}^+\mathrm{W}^-$}}

\begin{document}
\begin{titlepage}
\begin{center}
{\large   EUROPEAN LABORATORY FOR PARTICLE PHYSICS}
\end{center}
\bigskip
\begin{flushright}
CERN-PPE/97-168 \\ 
19 December, 1997
\end{flushright}
\bigskip\bigskip\bigskip\bigskip\bigskip
\begin{center}
{\huge \bf
Search for charged Higgs bosons 
in {\boldmath $\mathrm{e}^+\mathrm{e}^-$\unboldmath} collisions at
{\boldmath$\sqrt{s} = 130 - 172$\unboldmath} GeV}
\end{center}
\bigskip
\bigskip
\begin{center}{\LARGE The OPAL Collaboration}
%
% \bigskip\bigskip
%
% {\Large \bf Authors:}
%
% {\Large
% Beatrix Dienes, David Futyan, Dezs\H o Horv\'ath, P\'eter Ig\'o-Kemenes, 
% Gabriella P\'asztor, Graham Wilson, Terry Wyatt, Satoru Yamashita}
% 
%\bigskip
%\bigskip
% 
% {\Large \bf Editorial Board:}
% 
% {\Large
% Klaus Desch, Ehud Duchovni, Douglas Glenzinski, Richard Nisius}
%
\end{center}

\bigskip
\bigskip

\begin{center}
{\large \bf Abstract}
\end{center}
\noindent
A search is described to detect charged Higgs bosons via the process
$\ee\to\HH$, using data collected  by the OPAL detector  at  
center-of-mass energies of 130$-$172 GeV   with a total integrated
luminosity of 25 pb$^{-1}$. The decay channels 
are assumed  to be H$^+\to\qqp$ and H$^+\to\tau^+\nu_{\tau}$. 
No evidence for
charged Higgs boson production is observed. 
The lower limit for its mass is determined to be
52 GeV at 95\% confidence level, independent of the 
H$^+\to\tau^+\nu_\tau$ branching ratio.
%=========================================================
\bigskip\bigskip
\bigskip\bigskip
\bigskip
\bigskip
\begin{center}
{\large
(Submitted to Physics Letters B)} \\
\end{center}

\end{titlepage}
  
%======================== 
% Authors
%========================
\begin{center}{\Large        The OPAL Collaboration
}\end{center}\bigskip
\begin{center}{
%begin authorlist
K.\thinspace Ackerstaff$^{  8}$,
G.\thinspace Alexander$^{ 23}$,
J.\thinspace Allison$^{ 16}$,
N.\thinspace Altekamp$^{  5}$,
K.J.\thinspace Anderson$^{  9}$,
S.\thinspace Anderson$^{ 12}$,
S.\thinspace Arcelli$^{  2}$,
S.\thinspace Asai$^{ 24}$,
S.F.\thinspace Ashby$^{  1}$,
D.\thinspace Axen$^{ 29}$
% ...
%end authorlist
}\end{center}\bigskip
\bigskip
%begin institutes
$^{  1}$School of Physics and Astronomy, University of Birmingham,
Birmingham B15 2TT, UK
\newline
$^{  2}$Dipartimento di Fisica dell' Universit\`a di Bologna and INFN,
I-40126 Bologna, Italy
\newline
% ...
$^{  5}$Cavendish Laboratory, Cambridge CB3 0HE, UK
\newline
% ...
$^{  8}$CERN, European Organisation for Particle Physics,
CH-1211 Geneva 23, Switzerland
\newline
$^{  9}$Enrico Fermi Institute and Department of Physics,
University of Chicago, Chicago IL 60637, USA
\newline
% ...
$^{ 12}$Indiana University, Department of Physics,
Swain Hall West 117, Bloomington IN 47405, USA
\newline
% ...
$^{ 16}$Department of Physics, Schuster Laboratory, The University,
Manchester M13 9PL, UK
\newline
% ...
$^{ 23}$Department of Physics and Astronomy, Tel Aviv University,
Tel Aviv 69978, Israel
\newline
$^{ 24}$International Centre for Elementary Particle Physics and
Department of Physics, University of Tokyo, Tokyo 113, and
Kobe University, Kobe 657, Japan
\newline
% ...
$^{ 29}$University of British Columbia, Department of Physics,
Vancouver BC V6T 1Z1, Canada
\newline
% ...
$^{  e}$ and Department of Physics, New York University, NY 1003, USA
\newline
%end notes
 
\eject

%=======================================================================
\section{Introduction}
%=======================================================================

The interactions between elementary particles are well described by the
Standard Model (SM) \cite{sm} which assumes that  particle masses
are created via the  Higgs mechanism \cite{higgs}  through spontaneous
symmetry breaking.  The Standard Model contains one doublet
of complex scalar fields and predicts a single neutral Higgs boson.  
The minimal extension of the Higgs sector in the
Standard Model consists of two Higgs field doublets \cite{hunter} and predicts five
Higgs bosons of which three are neutral (h$^0$, H$^0$ and A$^0$) and
two are charged ($\Hp$ and $\Hm$). 
Despite a wide experimental effort, 
no evidence for Higgs bosons has yet been observed.
%
% There is no experimental evidence for the existence of a Higgs boson.

The discovery of a charged Higgs boson would be a clear indication 
of physics beyond the Standard Model.
Supersymmetry \cite{susy} is one of 
the possible extensions of the Standard Model.
The Minimal Supersymmetric Extension of the Standard Model (MSSM),
is the most popular example 
of such a model and contains two Higgs field doublets. 
At tree level it predicts 
that the charged Higgs boson is heavier than the \W\ boson, 
$\mH^2 = \mW^2 + {\mA}^2$.
Radiative corrections change this 
prediction~\cite{chmass}, however the detection of a charged Higgs 
boson lighter than the \W\ boson
would severely limit the parameter space of the MSSM. 

Searches for charged Higgs bosons were carried out at  $\Ecm\approx 91$ GeV
\cite{chlep1}, with  the  limit  $\mH\ > 44.1$ GeV being the most 
restrictive.
In 1995$-$96 the center-of-mass energy of the LEP collider has
been increased  in several steps up to 172 GeV. 
New lower bounds on the mass of the charged Higgs boson
above 50 GeV were recently reported~\cite{chlep2} by the
ALEPH and the DELPHI Collaborations.

On the basis of a study of the reaction b~$\to$ s$\gamma$,  the CLEO
Collaboration~\cite{cleoch} has set an indirect lower limit of $\mH\ >
(244 + 63/(\tan{\beta})^{1.3})$ GeV, where $\tan{\beta}$ is
the ratio of the  vacuum expectation values of the two Higgs fields. 
This limit is 
valid in the
two-doublet extensions of the Standard Model referred as
Model II, if the only new particles are the Higgs bosons.
However, in supersymmetric models, possible 
cancellations between contributions
of  the charged Higgs boson and  supersymmetric particles invalidate
this limit~\cite{goto}.

The CDF Collaboration recently reported a search for the decay t $\to$
bH$^+$ followed by H$^+ \to \tau^+\nu_\tau$. A lower limit of  
$\mH\ > 147 - 158$ GeV 
is derived for very large $\tan{\beta}$ depending on the
t$\bar{\mathrm t}$ production cross-section. For $\tan{\beta} < 40$ no
limit is derived, since in this regime the 
%
% top quark decays
%predominantly via t $\to$ bW$^+$
% 
assumed decay chain is no longer dominant \cite{cdfch}.

Charged Higgs bosons can be produced in pairs in the process $\ee \to
\HH$ with a  cross-section which to leading order depends only on the
Higgs boson mass and the center-of-mass energy~\cite{eehhcross}. The {\sc
Pythia}  program  \cite{pythia} is used  to calculate the charged Higgs
pair-production cross-section, including initial state
radiation, at the various \ee\ collision energies and for
various \Hpm\ masses. Higgs bosons decay 
predominantly to the heaviest fermions kinematically  allowed,  which
in the case of charged Higgs bosons can be
$\tau^+\nu_\tau$ or \csbar\ pairs, 
since the \cbbar\ decay mode is largely suppressed 
by the small CKM-matrix element, $\mathrm{V_{cb}}$. 

The branching ratio is model-dependent.  When
combining the results from the various
search channels BR$(\Hp\to\tpnu)+$BR$(\Hp\to\qqp)=1$ is assumed,
where BR$(\Hp\to\qqp)$ is the sum of all hadronic branching ratios
of the charged Higgs boson.

%=======================================================================
\section{OPAL detector and Monte Carlo generation}
%=======================================================================

The OPAL detector \cite{opal}, with its
acceptance of nearly $4\pi$ steradians, and its good tracking,
calorimetry and particle identification capabilities, is well suited
for this analysis which searches for widely different event topologies.
The apparatus is composed of a central tracking detector, consisting of a
silicon microvertex detector~\cite{si}
and several concentric  drift chambers
inside a 0.435~Tesla  magnetic field, surrounded by  presamplers, 
time-of-flight scintillators and a lead-glass electromagnetic
calorimeter located outside the magnet coil. The magnet return yoke is
instrumented for hadron calorimetry and is covered by external muon
chambers. Lead-scintillator detectors, the forward detector and gamma
catcher,  and silicon-tungsten calorimeters  close to the beam axis
complete the geometrical acceptance down to 25 mrad 
polar angle\footnote{The
OPAL coordinate system is a
right-handed 3-dimensional Cartesian coordinate system with its origin
at the nominal interaction point, $z$-axis along the nominal electron
beam direction and $x$-axis horizontal and directed towards the center of
LEP. The polar angle, $\theta$ is defined with respect to the +$z$ direction
and the azimuthal angle, $\phi$ with respect to the +$x$ direction.}.

The signal selection efficiencies and the background contributions  are
estimated  using Monte Carlo samples processed with a full simulation  
of the OPAL detector \cite{gopal}.  To generate the $\ee\to\HH$ events 
the {\sc Pythia}  \cite{pythia}  and {\sc Hzha} \cite{hzha}  program
packages are used. Both include initial and  final state radiation.
The generated partons are hadronized  using {\sc Jetset}
\cite{pythia}.  Signal samples of 500 events for the \csbar\cbars,
\tpnu\cbars\ and \tpnu\tmnu\ final states are produced at fixed values
of \mH\ between 40 and 75 GeV in steps of 5 GeV.  For systematic checks
some high statistics samples of 2500 events are also generated with
different quark flavors in the final state.

The background estimates from the different Standard Model processes are based on the
following event generators: 
{\sc Pythia} is used to generate  \qq($\gamma$) processes,
{\sc Excalibur} \cite{excalibur} and
{\sc Grc4f} \cite{grc4f} for four-fermion final states, 
{\sc Bhwide} \cite{bhwide} for  \ee$(\gamma)$, 
{\sc Koralz} \cite{koralz} for \mumu$(\gamma)$ and
\tautau$(\gamma)$, while {\sc Pythia}, 
{\sc Phojet} \cite{phojet}, {\sc Herwig}~\cite{herwig} and
{\sc Vermaseren}~\cite{vermaseren} for $\ee\qq$ and
${\mathrm e^+e^-}\ell^+\ell^-$ four-fermion final
states from two-photon processes.

%=======================================================================
\section{Event selection}
%=======================================================================

The present search is performed at  center-of-mass energies between 130
and 172 GeV, with  integrated luminosities measured by the
silicon-tungsten calorimeters of  approximately 2.5~pb$^{-1}$ at both
130 and 136 GeV, 10.0~pb$^{-1}$ at 161 GeV and
10.3 pb$^{-1}$ at 172 GeV with 0.5$-$1.4\% error, depending on the
center-of-mass energy, dominated by statistics.  
The analysis is sensitive to all dominant \HH\ final states, 
namely, the hadronic
\qqp\qppqppp, the semi-leptonic\footnote{The charge-conjugate final state
\qpq\tmnu\ is also implied.} \tpnu\qqp\ 
and the leptonic \tpnu\tmnu\ final states.
The integrated luminosities can differ from channel to
channel by less than 10\%, since
different detectors are required to be fully operational
in the different analyses. 

The event analysis uses charged particle tracks, electromagnetic and
hadronic calorimeter clusters selected by a set of quality
requirements similar to those used in previous Higgs
boson searches~\cite{smhpub}. The quality requirements applied in the 
search for the leptonic final state are described in 
Reference~\cite{ref-paper}.
Energy correction algorithms~\cite{gce,matching} are
used to prevent double counting in
the case of charged tracks and associated calorimeter clusters. 

%=======================================================================
\subsection{The leptonic final state}
%=======================================================================

A search for anomalous production of di-lepton events with
missing transverse momentum has been presented in
Reference~\cite{pr167} at \Ecm\ = 130$-$136 GeV and in 
Reference~\cite{ref-paper} at \Ecm\ = 161$-$172 GeV.
The latter includes a search for 
pair-produced charged Higgs bosons in  the
leptonic channel, $\HH\to\tpnu\tmnu$. 
At \sthree\ the results of the search for pair-produced
scalar tau leptons are used, since 
the experimental signature of 
$\HH\to\tpnu\tmnu$ is identical to that of scalar tau ($\tilde{\tau}^\pm$)
pair-production, 
$ \tilde{\tau}^+
\tilde{\tau}^-\to  \tau^+ \tilde{\chi}^0_1 \tau^- \tilde{\chi}^0_1$, 
for the case
when the lightest neutralino, $\tilde{\chi}^0_1$, is massless and stable.
The experimental methods and
results of the analyses are summarized below. For details refer
to~\cite{pr167,ref-paper}.

The signature for $\HH\to\tpnu\tmnu$  is a pair of tau leptons together
with missing energy and momentum. Tau leptons may be identified  by
their decays into electrons, muons  or hadrons. 
In selecting candidate events the missing
momentum is required to have a significant component 
in the plane perpendicular to the beam axis and 
the total missing momentum vector must point away from the beam axis.
Thereby Standard Model background with high energy particles
escaping down the beam pipe and giving rise 
to missing momentum along the beam axis, is rejected.
%
% A number of Standard Model
% processes can
% lead to high energy particles traveling down the beam pipe, thus being
% undetected and giving rise to missing momentum along the beam axis.
% Therefore, in selecting candidate signal events a significant missing
% momentum in the plane perpendicular to the beam axis is required  and 
% the total missing momentum vector must point away from the beam axis. 

A background that survives the above cuts arises from lepton
pairs  produced in two-photon processes
in which one of the initial state electron  is scattered at a
significant angle to the beam direction. Events that may have arisen
from such processes are suppressed by vetoing on energy being 
present in the forward detector, gamma catcher or silicon tungsten
calorimeters.

To further suppress the remaining Standard Model 
background mainly from W$^+$W$^-$ production and two-photon processes 
%
% sources of \llnunu\ events,
%
additional cuts on the momentum of the observed particles are applied. 
The cut values are optimized  separately for each value of \mH\ considered
using an automated optimization procedure. 

The results of the analysis are summarized in Table~\ref{tab:lepres}. 
Three candidates are selected in agreement with the Standard Model 
expectation. None of them have identified electrons or muons, 
so all tau lepton
candidates are consistent with hadronic decays.

\begin{table}[htb]
\begin{center}
\footnotesize
\begin{tabular}{||c||c|c|c|c|c|c|c|c||}
\hline\hline
$\roots$ & \multicolumn{8}{c||}{\mH\ (GeV)} \\
\cline{2-9}
 \makebox[7mm]{(GeV)} &  40 &  45 & 50 & 55 & 60  & 65 & 70 & 75 \\
\hline\hline
  \multicolumn{9}{||c||}{Number of events selected} \\
\hline
 133 
 & 0 & 0 & 0 & 0 & 0 
 &          0
 &          $-$
 &          $-$
 \\
 161
 &          1
 &          1
 &          1
 &          1
 &          1
 &          1
 &          1
 &          $-$
 \\
 172  
 &          2
 &          2
 &          2
 &          2
 &          2
 &          2
 &          2
 &          2
 \\
\hline\hline
\multicolumn{9}{||c||}{Number of events expected from 
Standard Model processes} \\
\hline
 133 
 & 1.1$\pm$0.4 & 1.1$\pm$0.4 & 1.1$\pm$0.4 & 1.1$\pm$0.4 & 1.1$\pm$0.4 
 & 1.1$\pm$0.4 &  $-$ &   $-$
 \\
 161
 &       0.9$\pm$0.2
 &       0.8$\pm$0.1
 &       1.0$\pm$0.2
 &       1.1$\pm$0.2
 &       1.1$\pm$0.2
 &       1.1$\pm$0.2
 &       1.0$\pm$0.2
 &          $-$
 \\
 172  
 &       1.0$\pm$0.1
 &       1.2$\pm$0.2
 &       1.3$\pm$0.2
 &       1.4$\pm$0.2
 &       1.4$\pm$0.2
 &       1.4$\pm$0.2
 &       1.4$\pm$0.2
 &       1.5$\pm$0.2
 \\
\hline\hline
  \multicolumn{9}{||c||}{Signal selection efficiency (\%)} \\
\hline
 133 
 & 27.6$\pm$1.5 & 31.1$\pm$1.5 & 34.6$\pm$1.5 & 37.6$\pm$1.5 & 38.2$\pm$1.5 
 & 40.2$\pm$1.6
 &          $-$
 &          $-$
 \\
 161
 & 44.2$\pm$2.2
 & 46.8$\pm$2.2
 & 50.2$\pm$2.2
 & 51.2$\pm$2.2
 & 53.0$\pm$2.2
 & 56.4$\pm$2.2
 & 59.0$\pm$2.2
 &          $-$
 \\
 172  
 & 29.0$\pm$2.0
 & 35.0$\pm$2.1
 & 40.4$\pm$2.2
 & 44.6$\pm$2.2
 & 46.2$\pm$2.2
 & 46.6$\pm$2.2
 & 51.2$\pm$2.2
 & 50.8$\pm$2.2
 \\
\hline\hline
\end{tabular}
\caption{
Leptonic Channel:  The number of selected  and  expected events  
together with selection efficiencies  at \mbox{$\protect\sqrt{s}=$} 
130$-$136, 161 and 172 GeV  for different values of \mH.
The errors are statistical only.
The dashes indicate masses which are kinematically forbidden or 
not simulated.
Note that 
there is significant overlap between the various \mH-dependent
selections.}
\label{tab:lepres}
\end{center}
\end{table}

In addition to the uncertainty due to the limited Monte Carlo statistics
a 5\%  systematic error is assigned to the
estimated selection efficiency to take into account deficiencies in the
Monte Carlo generators and the detector simulation. 

The dominant background  at $\sqrt{s}$ = 161$-$172 GeV results from
\wpair\ production which is well understood and the available  high
statistics Monte Carlo samples  describe well the OPAL data
\cite{ref:wwpapers}.  A 5\%  systematic error is assigned to the
estimated background to take into account the uncertainty in the
expected \wpair\ production cross-section arising from the uncertainty
in the W$^\pm$ boson 
mass and deficiencies in the Monte Carlo  detector simulation.
At \sthree\  the dominant background comes from   two-photon processes
which are less accurately modeled. The
expected  background  at this center-of-mass energy is conservatively set
to zero in the background subtraction procedure described in
Section~\ref{sec:results}.

%=======================================================================
\subsection{The semi-leptonic final state}
%=======================================================================

The  semi-leptonic channel $\HH\to\tpnu\qqp$ is characterized by  an
isolated tau lepton, a pair of acoplanar jets and sizeable missing
momentum due to the undetected neutrinos. 
The selection is described below.

\begin{itemize}

\item[(1)] The event must qualify as a hadronic final state
as defined in~\cite{line}.

\item[(2)]  There must be at least  one tau lepton identified following
Reference~\cite{smhlep2}, which has to be well isolated.
The flight direction of the tau lepton
is approximated by the direction of the momentum vector of its visible decay
products.  The ratio
of  both the track momenta ($R_{\mathrm tr}^{11/30}$) and 
the electromagnetic cluster energy ($R_{\mathrm em}^{11/30}$)
within an  $11^{\circ}$ half-angle cone relative 
to that within a $30^{\circ}$ 
half-angle cone around the direction of the tau lepton
should be larger than 0.95  and the cosine of
the angle between the direction of the tau lepton and 
the nearest  track should
be smaller than 0.94.  If there is more than one tau lepton in
the event and only one of them decays leptonically,  that one is kept,
otherwise the one with the largest $R_{\mathrm tr}^{11/30}$ is retained.

\item[(3)] Most of the two-photon and radiative two-fermion events are
eliminated by requiring that the polar angle of the missing momentum,
$\theta_p$, satisfies $|\cos{\theta_p}|<0.9$.

\item[(4)] Events with an energetic photon, identified
as electromagnetic cluster with energy greater than 15 GeV 
that has no track 
within a $30^\circ$ half-angle cone about the cluster axis, are rejected
to eliminate the remaining radiative events.

\item[(5)]  The two-fermion background is further reduced by
requiring the visible invariant mass of the event, $M_{\mathrm{vis}} =
\sqrt{E_{\mathrm{vis}}^2 -  \vec{P}_{\mathrm{vis}}^2}$, 
to be smaller than 0.8\,$\sqrt{s}$; 
the total missing momentum transverse to the beam direction,
$p_{\mathrm T}$, to be larger than 0.13\,$\sqrt{s}$ and
the aplanarity\footnote{
Aplanarity is defined as ${3\over 2}\lambda_3$, where $\lambda_i$ are the
eigenvalues [$\lambda_1\ge\lambda_2\ge\lambda_3$ with
$\lambda_1+\lambda_2+\lambda_3=1$] of the sphericity tensor
$S^{\alpha\beta}={\sum_{i}p_i^\alpha
p_i^\beta/\sum_{i}|{\bf p}_i|^2}$,
and measures the transverse momentum component out of the event plane.
}
to exceed 0.005.

\item[(6)]  At center-of-mass energies of 161$-$172 GeV 
to further suppress the
remaining four-fermion background, mainly from \WW\ with one
W$^\pm$ decaying leptonically, two additional conditions have to be satisfied. 
There should be no track in the event 
with momentum larger than 0.25\,$\sqrt{s}$ and 
the cosine of jet-jet angle in the hadronic system must 
exceed $-$0.65 ($-$0.55) at \Ecm\ = 161 (172) GeV. 
The two jets of the hadronic system 
obtained by removing the decay products of the tau lepton
are defined using the Durham jet-finding algorithm~\cite{durham}.
\end{itemize}

Table~\ref{tab:semileptoncuts} shows the  number
of selected data events, the total expected background 
 and the signal efficiency for \mH\ = 50 GeV after each cut at
all four center-of-mass energies. The agreement between data  and
background simulation is good. After all requirements 
no event is selected in the data
sample, while 2.7$\pm$0.2 (statistical error) events are expected
from Standard Model processes. Of these,
the four-fermion processes account for
13.3$\pm$3.0, 65.8$\pm$7.0 and 90.5$\pm$5.4\%  at 
\Ecm\ = 130$-$136, 161 and 172 GeV, respectively. 

\begin{table}[htb]
\centering
\begin{tabular}{||c||c|c|c||c|c|c||}
\hline\hline 
& \multicolumn{3}{|c||}{130 GeV} 
& \multicolumn{3}{|c||}{136 GeV} 
\\
\cline{2-7}
Cut& data  & \multicolumn{1}{|c|}{SM bgrd.} & Efficiency [\%] & 
data  & SM bgrd. & Efficiency [\%] \\ 
\hline\hline
1
&  736	& 728.8$\pm$3.2 	& 99.0$\pm$0.4
&  688	& 689.0$\pm$3.2 	& 97.8$\pm$0.7
\\
2
&   53	&  51.3$\pm$1.3 	& 61.0$\pm$2.2
&   56	&  47.2$\pm$1.3 	& 57.2$\pm$2.2
\\
3
& 28	& 25.0$\pm$0.8  	& 57.8$\pm$2.2
& 25	& 22.4$\pm$0.7  	& 54.8$\pm$2.2
\\
4 
& 26	& 21.5$\pm$0.7  	& 57.4$\pm$2.2
& 24	& 19.0$\pm$0.7  	& 54.8$\pm$2.2
\\
5 
&  0	& 0.5$\pm$0.1		& 40.4$\pm$2.2
&  0	& 0.2$\pm$0.1	   	& 42.4$\pm$2.2
\\
\hline\hline 
& \multicolumn{3}{|c||}{161 GeV}
& \multicolumn{3}{|c||}{172 GeV}\\
\cline{2-7}
Cut & data & SM bgrd.  & Efficiency [\%] & data & SM bgrd. & Efficiency [\%]\\ 
\hline\hline
1
& 1509	&1464.5$\pm$3.1 	& 98.6$\pm$0.5
& 1394	&1294.6$\pm$2.2 	& 98.2$\pm$0.6
\\
2
&  126	& 110.6$\pm$1.4   	& 67.4$\pm$2.1
&  131	& 116.3$\pm$1.1 	& 67.6$\pm$2.1
\\
3
& 46	& 48.8$\pm$0.7  	& 65.2$\pm$2.1
& 63	& 64.4$\pm$0.7	        & 65.8$\pm$2.1
\\
4 
& 41	& 42.9$\pm$0.7	        & 64.8$\pm$2.1
& 58	& 59.8$\pm$0.6  	& 64.8$\pm$2.1
\\
5 
&  2	& 4.1$\pm$0.1   	& 50.2$\pm$2.2
&  12	&15.4$\pm$0.2		& 44.8$\pm$2.2
\\
6 
&   0	&  0.7$\pm$0.1  	& 48.2$\pm$2.2
&   0	&  1.3$\pm$0.1  	& 44.2$\pm$2.2
\\
\hline\hline
\end{tabular}
\caption{
Semi-leptonic channel: Comparison of the number of observed events
and expected background  together with the selected
fraction of simulated signal  events (\mH\ = 50 GeV)  after each 
cut. The errors are statistical only.
}
\label{tab:semileptoncuts}
\end{table}

\begin{figure}[htbp] 
\centering
\epsfig{file=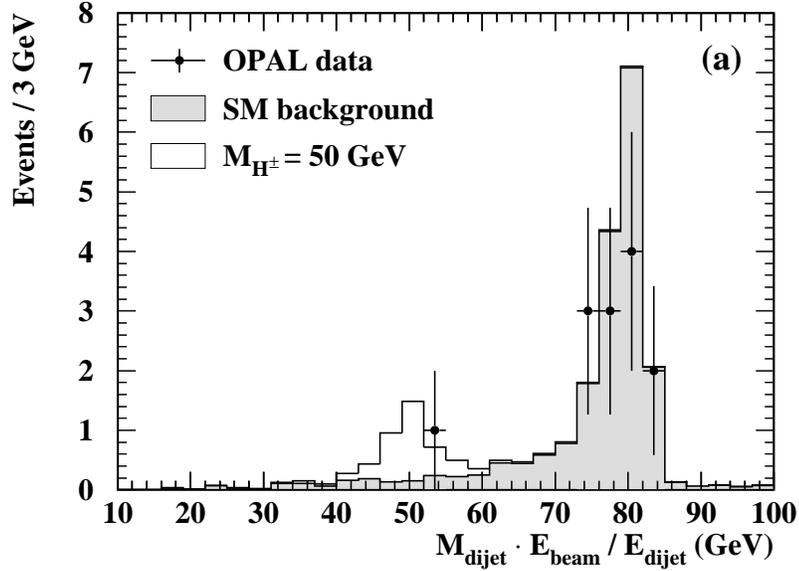,width=0.7\textwidth}
\epsfig{file=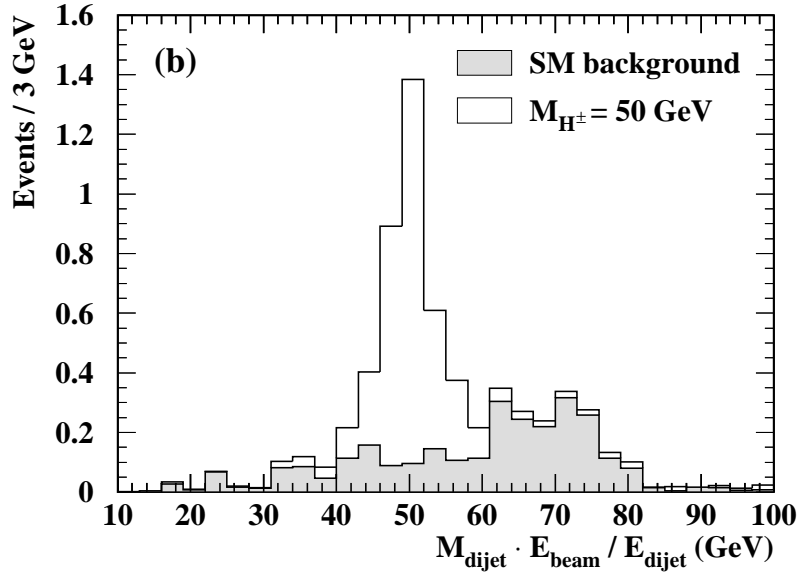,width=0.7\textwidth}
\caption{
Semi-leptonic channel: Scaled invariant mass distributions  at \Ecm\ =
130$-$172 GeV   normalized to the collected luminosity, (a) before cut
(6) and  (b) after all cuts. The selected
events are shown as dots with error bars,  the Standard Model background
estimate as a shaded histogram and a signal sample for \mH\ = 50 GeV
assuming  BR($\Hp\to\tpnu$) = 0.5 as an open histogram.
}
\label{fig:semilepres}
\end{figure}

In the semi-leptonic  channel the Higgs mass can be reconstructed from
the hadronic system  with 2$-$3 GeV resolution by  scaling the 
dijet invariant
mass by the ratio of the beam energy to the total
energy of the two jets. This  simple correction improves the mass
resolution by almost a factor of two and at the same time shifts the
mass of the \W\ bosons towards its nominal value, thereby decreasing
the expected background in the mass range below 65 GeV.  The
mass distributions  are shown in  Figure \ref{fig:semilepres} before
and after cut (6) for the selected events and the
expected background together with a signal of \mH\ = 50 GeV.
Note that cut (6) is also effective to reduce the background  
in the mass range below 60 GeV and 
that the remaining background is concentrated around a mass of 70 GeV.

The flavor independence of the selection
is tested  using Monte Carlo samples of  $\HH\to\ \tpnu\cbars$ and
$\HH\to\tpnu{\mathrm b}\bar{\mathrm c}$. The  observed differences are
consistent within the statistical  error of $2.4\%$, which is
conservatively incorporated into the systematic error.

\enlargethispage{-12pt}

The signal selection efficiencies are affected by the following
uncertainties: Monte Carlo statistics, see Table~\ref{tab:semilepton};
uncertainties on the tau lepton identification efficiency (including the
errors on electron and muon identification), 3\%;  modeling of the
cut variables excluding the tau lepton identification, 6\%; and dependence
on  the flavor of the final state quarks, 2.4\%. 
%
% The uncertainties in 
% modeling  the selection variables are estimated by displacing each cut
% value by an amount corresponding to the difference of the mean values
% of the data and background Monte Carlo distributions.

\begin{table}[htb]
\begin{center}
\small
\begin{tabular}{||c||c|c|c|c|c|c|c|c||}
\hline\hline 
$\sqrt{s}$ & 
\multicolumn{8}{|c||}{Signal selection efficiencies ($\%$) for \mH\  } 
%& \multicolumn{2}{|c||}{Events}
\\ \cline{2-9}
% \cline{7-8}
(GeV) & 40 GeV  & 45 GeV & 50 GeV & 55 GeV & 60 GeV 
& 65 GeV & 70 GeV & 75 GeV  \\ 
\hline\hline
130 & 
% off & 
37.0$\pm$2.2 	& 42.4$\pm$2.2
&40.4$\pm$2.2	& 35.0$\pm$2.1
&27.6$\pm$2.0 	& $-$
&$-$& $-$
\\ 
% \hline 
136 &
% off & 
37.0$\pm$2.2	& 45.6$\pm$2.2
&42.4$\pm$2.2 	& 38.6$\pm$2.2 
&35.0$\pm$2.1 	& $-$
&$-$& $-$
\\  
% \hline 
%161 &
% off &
%\\ 
161 &
% on & 
  42.6$\pm$2.2 &  46.0$\pm$2.2 
& 48.2$\pm$2.2 &  43.0$\pm$2.2 
& 35.0$\pm$2.1 & 31.0$\pm$2.1 
&26.8$\pm$2.0& $-$
\\  
% \hline 
%172  & off &
%\\ 
172 &
% on & 
  41.8$\pm$2.2 & 43.2$\pm$2.2  
& 44.2$\pm$2.2 & 42.2$\pm$2.2  
& 37.0$\pm$2.2 & 35.8$\pm$2.1
& 29.8$\pm$2.0 & 12.2$\pm$1.5
\\ 
\hline \hline 
\end{tabular}
\end{center} 
\caption{Semi-leptonic channel: Signal selection efficiencies for the
various center-of-mass energies  and charged Higgs masses.   The
errors are statistical only.
The dashes indicate masses which are kinematically forbidden or 
not simulated.
For higher masses the selection efficiency
drops due to cut (6).
} 
\label{tab:semilepton}
\end{table} 

The background estimate has the following errors:  Monte Carlo
statistics, see Table~\ref{tab:semileptoncuts};
modeling of the hadronization process estimated by comparing different event
generators, 9\%; modeling of the variables used to identify tau leptons, 
5\%; and modeling of the remaining selection variables, 5\%.

\eject

%=======================================================================
\subsection{The hadronic final state}
%=======================================================================
 
The hadronic channel, $\HH \to \qqp \qppqppp$,  is
characterized by an event topology with four well separated hadron
jets and large visible energy. The selection is described
below.

\begin{itemize}

\item[(1)] 
The event must qualify as hadronic final state as defined in~\cite{line}.

\item[(2)]  Events with a radiative photon or large missing energy are
eliminated by requiring
the effective center-of-mass energy, $\sqrt{s^\prime}$,
calculated as described in Reference~\cite{sprim}, 
to be at least $0.87\sqrt{s}$ and 
the visible invariant mass to be at least $0.7\sqrt{s}$.

\item[(3)] The events are reconstructed into four jets using the Durham
jet-finding
algorithm~\cite{durham} with the visible energy as the scale parameter.
The jet resolution parameter, $y_{34}$, at which the number of jets
changes from 3 to 4, has to be larger than 0.01 at \Ecm\ =
130$-$136 GeV, and larger than 0.005 at \Ecm\ = 161$-$172 GeV. The
tighter cut at lower energies is necessary because of the higher
\qq\ background.
Moreover each jet must contain at least one charged track.   

\item[(4)] At \Ecm\ = 161$-$172 GeV, the remaining radiative 
$\qq\gamma$  events are further suppressed by vetoing on
jets with properties
compatible with those of a radiative photon, namely, exactly one
electromagnetic cluster, not more than two tracks  
and jet energy above $\sqrt{s}-121$ GeV. 

\item[(5)]
To further reduce the \qq\ background  the following
requirements are imposed:
the polar angle of the thrust axis has to satisfy
$|\cos\theta_{\rm thr}| < 0.8$;
the event shape parameter\footnote{ The $C$ parameter is
defined as $C=(\lambda_1\lambda_2+
\lambda_1\lambda_3+\lambda_2\lambda_3)$, where $\lambda_i$ are the
eigenvalues
% [$\lambda_1\ge\lambda_2\ge\lambda_3$ with 
[$\lambda_1+\lambda_2+\lambda_3=1$] of the generalized sphericity tensor
$S^{(1)\alpha\beta}={\sum_{i}(p_i^\alpha p_i^\beta/|{\bf
p}_i|)/\sum_{i}|{\bf p}_i|}$.}, $C$, 
has to be larger than 0.6 at \Ecm\ = 130$-$136 GeV
and larger than 0.45 at \Ecm\ = 161$-$172 GeV; and
the cosine of the angle between any pair of jets 
must be smaller than 0.62 at \sthree\ and 0.66 at higher
center-of-mass energies.

\item[(6)]  To test the compatibility of the event 
with the decay of two equal mass objects
a four-constraint kinematic fit  
requiring energy and momentum conservation is performed and
the mass difference between the two dijet systems is 
calculated for all three possible jet pair combinations.
The event is  discarded if the $\chi^2$-probability of the fit is below 0.01
or if the smallest mass difference is larger 
than 6 GeV at \Ecm\ = 130$-$136 GeV and 
8 GeV at higher center-of-mass energies.
For all events passing this cut, to obtain the
best possible dijet mass resolution
a five-constraint kinematic fit is performed 
for all three jet pair combinations
imposing energy and momentum conservation and equal dijet invariant
masses and the  event is rejected 
if the largest $\chi^2$-probability is below 0.01.

\item[(7)] At center-of-mass energies of 161 GeV and
above a veto is applied against \WW\
events using the dijet masses calculated after the four-constraint
kinematic fit.  
At \Ecm\ = 161 GeV, since the \W\ bosons are produced practically at rest,
the two jets having the largest measured opening
angle are assigned to one of the \W\ bosons and the two remaining jets
to the other. An event is
rejected if both jet pairs have an invariant mass greater than 70 GeV.
At \sseven\ the
event is rejected if any of the three possible jet pair combinations
yields invariant masses greater than 74 GeV  for  both of the two dijet
systems.

\end{itemize}

For the remaining events, the jet pair association giving 
the highest $\chi^2$-probability in the five-constraint kinematic fit
is retained.
The resulting mass resolution is  1.0$-$1.5 GeV.

\begin{figure}[htbp]
\centering
% \begin{center}
\epsfig{file=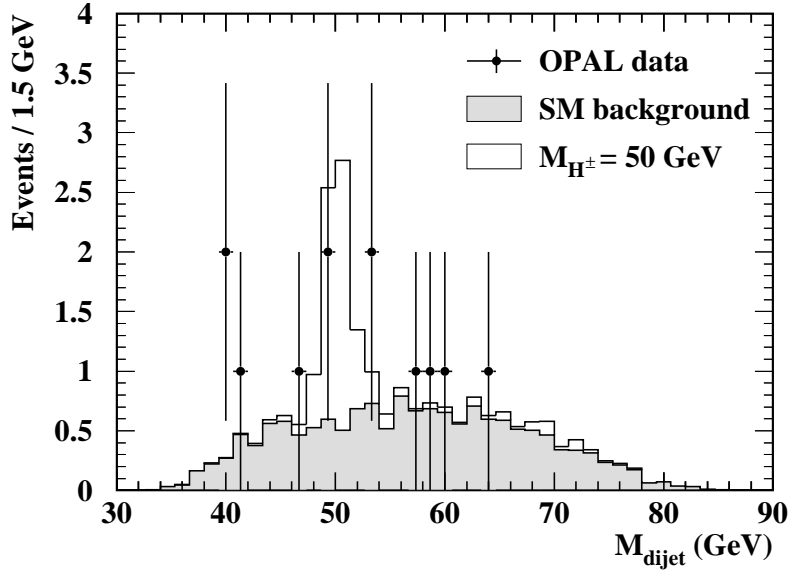,width=0.7\textwidth}
% \end{center}
\caption{ Hadronic channel:
Invariant mass distribution using a 
five-constraint  kinematic fit  at
\Ecm\ = 130$-$172 GeV normalized to the 
collected luminosity, after all
cuts.
The selected events are shown as dots with error bars, 
the Standard Model background estimate as a shaded histogram and
a signal sample for \mH\ = 50 GeV
assuming BR($\Hp\to\qqp$) = 1 as an open histogram.  
}
\label{fig:hadmres}
\end{figure}

Table~\ref{tab:hadcuts} shows the  number of selected events, the
 estimated background and the fraction of signal events retained for \mH\ =
50 GeV at all center-of-mass energies after each cut. The
agreement between data and expected background is good. In total, twelve events
are selected in the data, while 15.3$\pm$0.4 (statistical
error) events are expected from Standard Model processes. 
The four-fermion processes account for 10.1$\pm$1.2,
32.3$\pm$1.7 and 64.8$\pm$3.5\% of
the expected background at
\Ecm\ = 130$-$136, 161 and 172 GeV, respectively. Figure
\ref{fig:hadmres} shows the invariant mass distribution of the selected
events together with the Standard Model background expectation and a signal of \mH\ = 50
GeV.

\begin{table}[htp]
\centering
\begin{tabular}{||c||c|c|c||c|c|c||}
\hline\hline 
& \multicolumn{3}{|c||}{130 GeV} 
& \multicolumn{3}{|c||}{136 GeV} 
\\
\cline{2-7}
Cut & data  &  SM bgrd. & Efficiency [\%] & data & SM bgrd. & Efficiency [\%]\\ 
\hline\hline
1  &	744& 	733.8$\pm$3.2	&99.8$\pm$0.2&
	676&	679.7$\pm$3.1	&99.8$\pm$0.2  \\
% \hline
2  &	173& 	201.7$\pm$2.1&	94.2$\pm$1.0&
	184&	180.2$\pm$1.9&	93.0$\pm$1.1 \\
% \hline
3  &	11& 	11.5$\pm$0.6&	62.8$\pm$2.2&
	14&	11.6$\pm$0.6&	59.4$\pm$2.2 \\
% \hline
% 4  &$-$&	$-$&	$-$&	$-$&	$-$&	$-$ \\
% \hline
5  & 	4&	3.7$\pm$0.3&	49.2$\pm$2.2&
	4&	3.9$\pm$0.3&	50.4$\pm$2.2  \\
% \hline
6  & 	2&	1.2$\pm$0.2&	33.4$\pm$2.1&
	2&	1.6$\pm$0.2&	34.4$\pm$2.1\\
% \hline
% 8  & $-$&	$-$&	$-$&	$-$&	$-$&	$-$\\
\hline
\hline 
& \multicolumn{3}{|c||}{161 GeV}
& \multicolumn{3}{|c||}{172 GeV}\\
% \cline{2-3} \cline{4-5} \cline{6-8}
\cline{2-7}
Cut & data &  SM bgrd.  & Efficiency [\%]  & data & SM bgrd.& Efficiency [\%]\\ 
\hline\hline
1  &	1497 & 	1453.5$\pm$3.1	&	100.0$\pm$0.2 & 
 	1393 &	1310.8$\pm$3.4	&	100.0$\pm$0.2 \\
% \hline
2  & 	392&	374.5$\pm$1.4		&	93.6$\pm$1.1&
	359&	368.9$\pm$1.3		&	94.6$\pm$1.0 \\
% \hline
3  &	62 &	 53.2$\pm$0.5   	&	75.6$\pm$1.9 &
	88 &	 82.5$\pm$0.5		&	72.2$\pm$2.0\\
% \hline
4  & 	59 &	50.8$\pm$0.5		&	75.6$\pm$1.9  &
	87 & 	80.2$\pm$0.5		&	72.2$\pm$2.0 \\
% \hline
5  & 	21&	19.2$\pm$0.3		&	67.2$\pm$2.1&
	36&	38.1$\pm$0.3		&	59.4$\pm$2.2\\
% \hline
6  &	8& 	8.9$\pm$0.2		&	52.4$\pm$2.2&
	14&	18.6$\pm$0.3		&	48.8$\pm$2.2\\
% \hline
7  &	3 &	5.0$\pm$0.2		&	45.4$\pm$2.2 &
	5 &	7.5$\pm$0.2		&	42.2$\pm$2.2 \\
\hline\hline
\end{tabular}
%\end{center}
\caption{Hadronic channel:
Comparison of the number of observed events and expected
background together with the selected fraction of simulated
signal  events (\mH\ = 50 GeV) after each cut. 
The errors are statistical only.
}
\label{tab:hadcuts}
\end{table}

The  systematic effects on the signal selection efficiency are the
following:  Monte Carlo statistics, see Table~\ref{tab:hadr};
final-state quark flavor dependence, 2.4\%; and modeling of the
cut variables, 5\%.

%\begin{sidewaystable}[p]
\begin{table}[htp]
\begin{center}
\small
%\footnotesize
\begin{tabular}{||c||c|c|c|c|c|c|c|c||}
\hline\hline 
$\sqrt{s}$ & 
%W veto &
\multicolumn{8}{|c||}{Signal selection efficiencies ($\%$) for \mH\ }\\
%& \multicolumn{2}{|c||}{Events}
%\\ \cline{3-7} \cline{8-9}
\cline{2-9}
% \cline{7-8}
(GeV) 
%&
& 40 GeV  & 45 GeV & 50 GeV & 55 GeV & 60 GeV 
%& SM bgrd. & data \\ 
& 65 GeV & 70 GeV & 75 GeV \\
\hline\hline
130
% &
& 32.4$\pm$2.1 & 32.4$\pm$2.1 & 33.4$\pm$2.1 & 29.6$\pm$2.0 & 22.2$\pm$1.9 
%& 1.27$\pm$0.27 & 2
&$-$&$-$& $-$
\\ 
% \hline 
136
% &
& 29.8$\pm$2.0 & 36.6$\pm$2.2 & 34.6$\pm$2.1 & 26.0$\pm$2.0 & 26.8$\pm$2.0 
% & 1.49$\pm$0.26 & 2
&$-$&$-$& $-$
\\ 
% \hline 
% 161 & off
% & 41.6$\pm$2.2 & 47.0$\pm$2.2 & 53.2$\pm$2.2 & 47.2$\pm$2.2 & 40.6$\pm$2.2 
% & 8.71$\pm$0.18 & 8
%\\ 
161 
%& on
& 36.0$\pm$2.1 & 41.4$\pm$2.2 & 45.4$\pm$2.2 & 41.4$\pm$2.2 & 36.4$\pm$2.2 
%& 5.19$\pm$0.16 & 3
& 31.4$\pm$2.1 & 28.0$\pm$2.0 & $-$
\\
% \hline 
% 172  & off
% & 28.2$\pm$2.0 & 48.0$\pm$2.2 & 49.6$\pm$2.2 & 49.0$\pm$2.2 & 42.6$\pm$2.2 
% & 17.93$\pm$0.19 & 15\\ 
172 
%& on
& 24.6$\pm$1.9 & 40.4$\pm$2.2 & 42.2$\pm$2.2 & 39.6$\pm$2.2 & 39.0$\pm$2.2 
%& 7.19$\pm$0.15 & 5
&29.4$\pm$2.0&31.2$\pm$2.1&20.0$\pm$1.8
\\ \hline \hline 
\end{tabular}

\end{center}
\caption{ Hadronic channel:
Signal selection efficiencies 
for the various center-of-mass energies 
and charged Higgs masses. 
The errors are statistical only. 
The dashes indicate masses which are kinematically forbidden or 
not simulated.
For higher masses the detection
efficiency drops due to cut (7).
}
\label{tab:hadr}
\end{table}
%\end{sidewaystable}

The background estimate is affected by the following systematic
uncertainties: limited Monte Carlo statistics, see Table~\ref{tab:hadcuts}; 
 modeling of the hadronization process estimated
by comparing different event generators, 7\%;  modeling
of the cut variables, 6\%. Since
the theoretical uncertainty on the prediction of the QCD four-jet rates is
not known,  conservatively its experimental error of 15\% \cite{QCD4j}
is taken which is dominated by statistics. Taking into account
the relative weight of the QCD background, this results in an 8\% error on the
background estimate.

%=======================================================================
\section{Results} \label{sec:results}
%=======================================================================

The statistical method of Reference \cite{bock} is used to calculate 95\%
confidence level lower limits on the charged Higgs boson mass. This
method has been developed to derive exclusion limits for particle
searches when several candidate events are observed in different decay
channels with different mass resolutions and different background
conditions. The method
introduces an event weight
%
% \footnote{
% The event weight for a channel $i$ is inversely proportional to 
% $1+b_i\Sigma_j s_j/s_i$, where $b_i$ is the background level
% and $s_i$ is the expected number of signal events, differential in mass.} 
%
for each channel 
 and derives the confidence limit from the sum of the
event weights for all candidates. The mass spectrum of the
background is also taken into account. 

The predicted background is accounted for by considering the selected
events as {\sl signal plus background}. In the calculation
of the limit the expected background is decreased by its
statistical and systematic error.

\begin{figure}[htbp]
\centering
% \begin{center}
\null
\vspace{-1.3cm}
\null
\epsfig{file=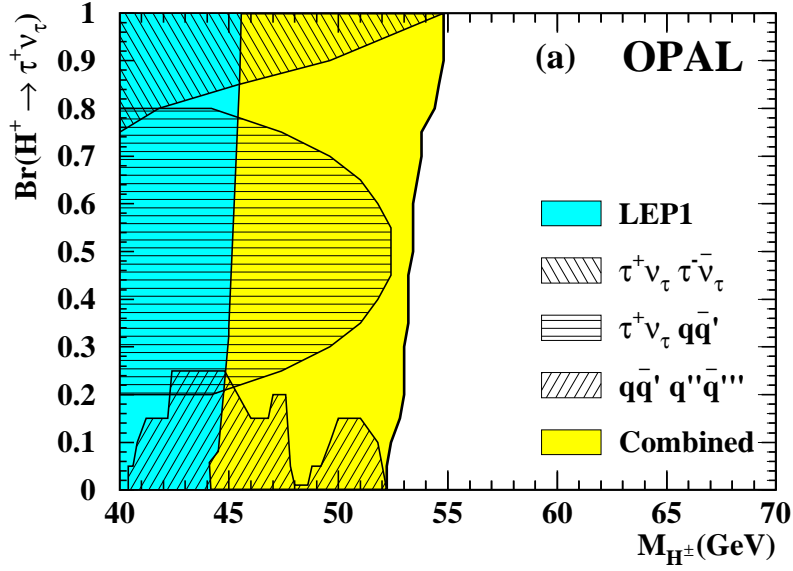,width=0.7\textwidth}
% \null
% \vspace{-1cm}
% \null
\epsfig{file=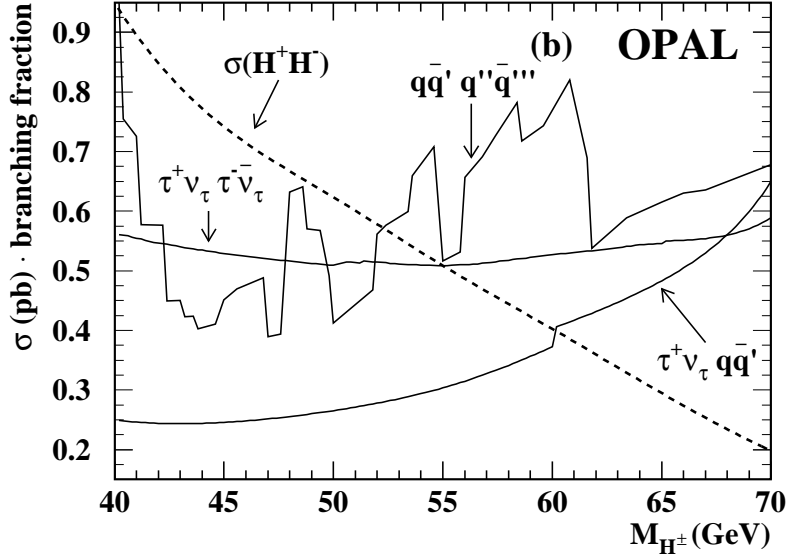,width=0.7\textwidth}
% \vspace{-0.2cm}
% \end{center}
\caption{(a) Excluded areas at 95\% confidence level
 in the $[M_{{\rm H}^\pm}$, 
BR$(\Hp\to\tpnu)]$ plane. The dark shaded area is excluded by  LEP1 and
the light shaded area by the present  search. The 
results from each of the channels separately are indicated by
different hatch styles.
(b) Upper limits, scaled to  \Ecm\ = 172 GeV,  on the  production
cross-section times branching fraction of the decay to  a given
final state, at 95\% confidence level, calculated using the $s$-dependence of the 
charged Higgs boson production cross-section for each channel. 
The charged Higgs boson production cross-section 
at \Ecm\ = 172 GeV is also shown as a dashed line. Note that the maximum 
branching fraction for the \tpnu\qqp\ final state is 0.5.
}
\label{fig:xsec}
\end{figure}

The lower bounds on the mass of the charged Higgs boson, at the 95\%
confidence level, obtained from the searches in the leptonic, semi-leptonic and
hadronic channels, are presented in Figure~\ref{fig:xsec} as a function
of the $\Hp\to\tpnu$ branching ratio. The limits are obtained using
the cross section calculated by {\sc Pythia} for the process
$\ee\to\HH$. They  take into account the integrated
luminosities of the data and the selection efficiencies
as a function of $\mH$ at each center-of-mass energy. 
The uncertainty on the signal efficiency 
is incorporated into the limit using the method
described in Reference~\cite{ref:cousin}.

Charged Higgs bosons are excluded
at  95\% confidence level
 independent of the $\Hp\to\tpnu$ branching ratio  up
to a mass of 52 GeV. 
%
% Mass limits are given in
% Table~\ref{tab:excl} for selected values of 
% the $\Hp\to\tpnu$ branching ratio. 
%
% \begin{table}[htp]
% \begin{center}
% \begin{tabular}{||l||c|c|c|c|c|c|c|c||} \hline\hline
% BR$(\Hp\to\tpnu)$ & 0.0  & 0.25 & 0.50 & 0.75 &  1.0 \\ \hline
% $\mH$ [GeV]& 52.0 & 53.0 & 53.4 & 53.8  & 54.8  \\ \hline
% \hline
% \end{tabular}
% \end{center}
% \caption[sig]{The 95\% CL lower limits on $\mH$ for selected 
% values of the $\Hp\to\tpnu$ branching ratio.
% }
% \label{tab:excl}
% \end{table}
%
Upper limits on the production cross-section 
times branching fraction of the decay to a given final state
assuming the $s$-dependence of the charged Higgs boson production
cross-section, 
scaled to \Ecm\ = 172 GeV,
are presented in Figure \ref{fig:xsec} for all three 
final states.

%==================
% Acknowledgements
%==================

\section*{Acknowledgements}

We particularly wish to thank the SL Division for the efficient operation
of the LEP accelerator at all energies
 and for
their continuing close cooperation with
our experimental group.  We thank our colleagues from CEA, DAPNIA/SPP,
CE-Saclay for their efforts over the years on the time-of-flight and trigger
systems which we continue to use.  In addition to the support staff at our own
institutions we are pleased to acknowledge the  \\
Department of Energy, USA, \\
National Science Foundation, USA, \\
Particle Physics and Astronomy Research Council, UK, \\
Natural Sciences and Engineering Research Council, Canada, \\
Israel Science Foundation, administered by the Israel
Academy of Science and Humanities, \\
Minerva Gesellschaft, \\
Benoziyo Center for High Energy Physics,\\
Japanese Ministry of Education, Science and Culture (the
Monbusho) and a grant under the Monbusho International
Science Research Program,\\
German Israeli Bi-national Science Foundation (GIF), \\
Bundesministerium f\"ur Bildung, Wissenschaft,
Forschung und Technologie, Germany, \\
National Research Council of Canada, \\
Research Corporation, USA,\\
Hungarian Foundation for Scientific Research, OTKA T-016660, 
T-023793 and OTKA F-023259.\\

%=======================================================================
% References
%=======================================================================
%\newpage

\end{document}